\documentclass[%
 aip,
 amsmath,amssymb,
 reprint,%
 author-year,%
 floatfix
]{revtex4-1}

\usepackage{graphicx}
\usepackage{dcolumn}
\usepackage{bm}

\usepackage[utf8]{inputenc}
\usepackage[T1]{fontenc}
\usepackage{mathptmx}

\usepackage{hyperref}
\PassOptionsToPackage{hidelinks}{hyperref} 
\PassOptionsToPackage{final}{hyperref}

\DeclareTextFontCommand{\mytexttt}{\ttfamily\hyphenchar\font=45\relax}

\usepackage{listings}
\usepackage{xcolor}

\definecolor{codegreen}{rgb}{0,0.6,0}
\definecolor{codegray}{rgb}{0.5,0.5,0.5}
\definecolor{codepurple}{rgb}{0.58,0,0.82}
\definecolor{backcolour}{rgb}{0.975,0.975,0.975}
 
\lstdefinestyle{mystyle}{
    backgroundcolor=\color{backcolour},   
    commentstyle=\color{codegreen},
    keywordstyle=\color{magenta},
    numberstyle=\tiny\color{codegray},
    stringstyle=\color{codepurple},
    basicstyle=\ttfamily\scriptsize,
    breakatwhitespace=true,         
    breaklines=true,                 
    captionpos=b,                    
    keepspaces=false,                 
    numbersep=5pt,                  
    showspaces=false,                
    showstringspaces=false,
    showtabs=false,                  
    tabsize=2,
    frame=bt,
    rulecolor=\color{gray},
    language=Python,
}

\usepackage{hyphenat}

\begin{document}

\preprint{AIP/123-QED}

\title[Magnetic-field modeling with surface currents:  Implementation and usage of bfieldtools]{Magnetic-field modeling with surface currents:  Implementation and usage of bfieldtools}

\author{Rasmus Zetter}
 \email{rasmus.zetter@aalto.fi}
 \affiliation{Department of Neuroscience and Biomedical Engineering, Aalto University School of Science, FI-00076 Aalto, Finland}
 
\author{Antti J Mäkinen}
\affiliation{Department of Neuroscience and Biomedical Engineering, Aalto University School of Science, FI-00076 Aalto, Finland}

\author{Joonas Iivanainen}
\affiliation{Department of Neuroscience and Biomedical Engineering, Aalto University School of Science, FI-00076 Aalto, Finland}

\author{Koos C J Zevenhoven}
\affiliation{Department of Neuroscience and Biomedical Engineering, Aalto University School of Science, FI-00076 Aalto, Finland}

\author{Risto J Ilmoniemi}
\affiliation{Department of Neuroscience and Biomedical Engineering, Aalto University School of Science, FI-00076 Aalto, Finland}

\author{Lauri Parkkonen}
\affiliation{Department of Neuroscience and Biomedical Engineering, Aalto University School of Science, FI-00076 Aalto, Finland}

\date{\today}

\begin{abstract}
We present a novel open-source Python software package, \mytexttt{bfieldtools}, for magneto-quasistatic calculations with current densities on surfaces of arbitrary shape. The core functionality of the software relies on a stream-function representation of surface-current density and its discretization on a triangle mesh. Although this stream-function technique is well-known in certain fields, to date the related software implementations have not been published or have been limited to specific applications. With \mytexttt{bfieldtools}, we aimed to produce a general, easy-to-use and well-documented open-source software. The software package is written purely in Python; instead of explicitly using lower-level languages, we address computational bottlenecks through extensive vectorization and use of the NumPy library. The package enables easy deployment, rapid code development and facilitates application of the software to practical problems. In this paper, we describe the software package and give an extensive demonstration of its use with an emphasis on one of its main applications -- coil design.
\end{abstract}

\maketitle

\section{Introduction}

\noindent Within many fields of engineering and science, there is a need for modeling the relationship between magnetic fields and surface currents in complex geometries. For example, to model eddy currents in conducting sheets, one needs to calculate the coupling between the external field and the currents as well as the inductive effects of the currents within the conductor \citep[e.g.][]{peeren_stream_2003, zevenhoven_dynamical_2015}. Such modeling is also useful in computing the magnetic noise arising from thermal fluctuations \citep{roth_thermal_1998, sandin_noise_2011, iivanainen_general_2020} and in designing surface-current patterns that generate a desired magnetic field. Finally, through such field calculations, surface currents can be used as equivalent sources in reconstruction and interpolation of magnetic fields, e.g., in geo- \citep{mendonca_equivalent_1994, blakely_potential_1996} and biomagnetism \citep{taulu_presentation_2005}.

A current density is often represented using a set of basis functions. For currents on simple domains (such as planes, cylinders, toroids, or spheres), basis functions can be formed analytically \citep[e.g.][]{turner_target_1986, merkel_solution_1987, crozier_design_1995, liu_spherical_1997, drevlak_automated_1998, roth_thermal_1998, suits_optimizing_2003, forbes_novel_2004, zevenhoven_conductive_2014}.  \citet{pissanetzky_minimum_1992} introduced a general stream-function representation of the surface-current density on arbitrary surfaces, which discretizes the current on triangle surface meshes in a manner similar to finite-element and boundary-element methods (FEM and BEM, respectively).

Within the field of magnetic resonance imaging (MRI), triangle mesh -based stream-function methods have been applied to magnetic field modeling and coil design since the early 1990s \citep[e.g.][]{pissanetzky_minimum_1992, peeren_stream_2003, lemdiasov_stream_2005, poole_improved_2007, harris2013shielded}. Similar methods have also been used in plasma physics \citep{abe_new_2003}. More recently, the same principles have been used in the design of coils for transcranial magnetic stimulation (TMS) \citep{koponen_coil_2017,cobos_sanchez_inverse_2018} as well as magnetic nanoparticle imaging \citep{bringout_performance_2015}.

Still, these coil-design techniques and surface-current models have most often been applied to simple geometries and their implementations have not been available for wider audiences. While the basic equations or concepts may be well-known, implementing, testing and validating such software requires considerable time and effort, something that may not be available for all prospective users.

In this paper, we present a novel open-source Python software package for magnetic field modeling and coil design, \mytexttt{bfieldtools} (available at \url{https://bfieldtools.github.io}). This paper focuses on describing the software package itself and demonstrates its usage through several examples. While we give a brief overview of the working principles behind the software in the following section, for a more thorough treatment of the underlying physics and computational aspects we refer to our accompanying publication \citep{makinen_magnetic-field_2020}. 

\section{Computations using the discrete stream function}
\label{sec:stream}
\noindent \mytexttt{bfieldtools} uses the scalar stream-function representation of a surface current density \citep{pissanetzky_minimum_1992, peeren_stream_2003}, which is discretized as a piecewise linear function onto a triangle mesh. Compared to analytical methods that require certain symmetries for the source-current distributions, the use of triangle meshes as source domains provides the user with considerable geometrical freedom.

\begin{figure}[t]
    \centering
    \includegraphics[width=\columnwidth]{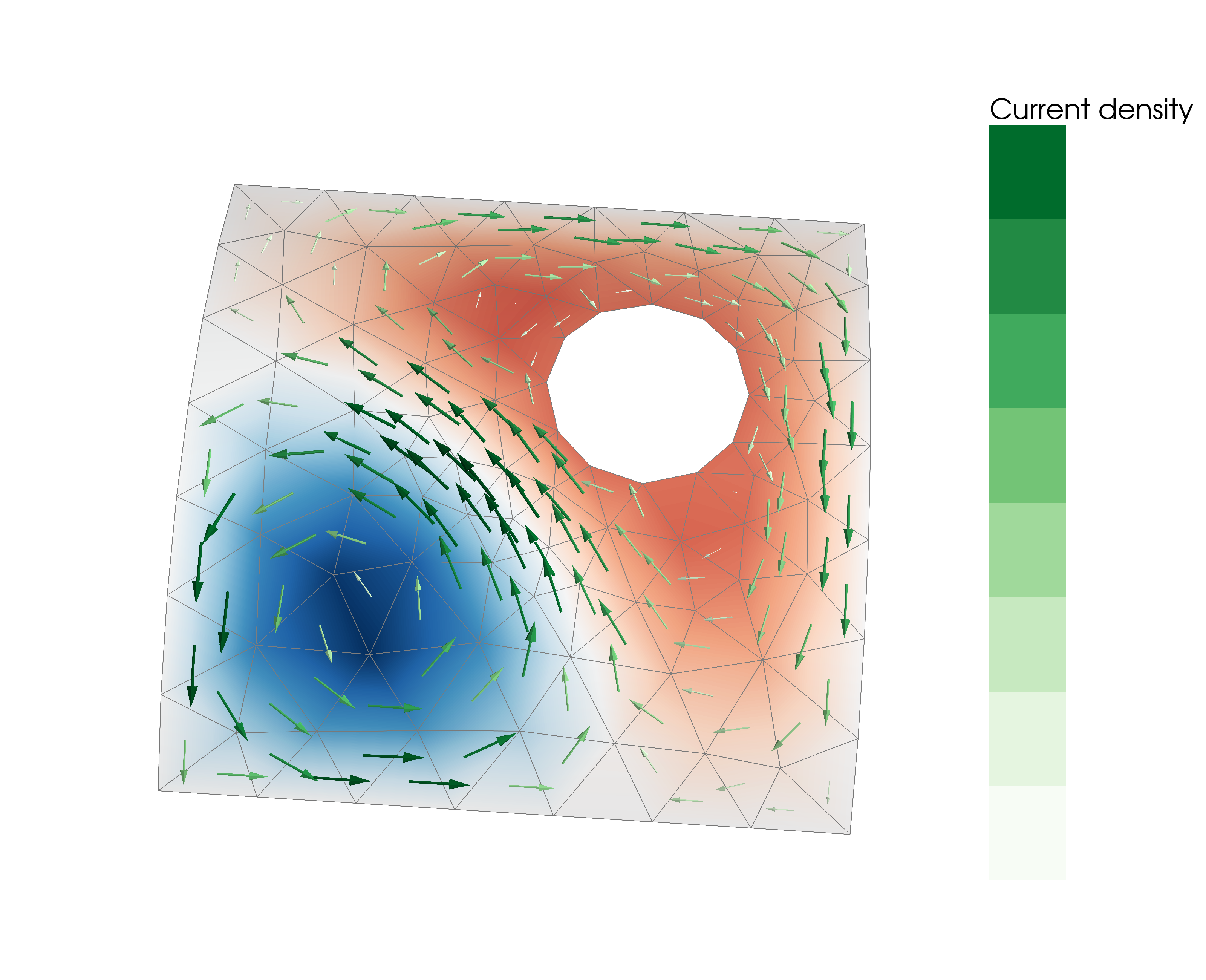}
    \caption{An example stream function (red--blue colormap) and its rotated gradient, i.e. the surface current density (arrows; green colormap) on a surface mesh with a hole in it. The surface normal is oriented up towards the reader.}
    \label{fig:streamfunction_gradient}
\end{figure}

The triangle mesh discretization is based on approximating the stream functions linearly on the face of each triangle as in finite-element methods (FEM), and as illustrated in Fig. \ref{fig:streamfunction_gradient}. A piecewise linear stream function is defined on the surface using so-called hat functions, which are defined as having the value one at a single vertex and falling linearly to zero at the edges of the triangles neighboring the vertex. The stream function $\psi$ can then be represented as a linear combination of the hat functions $h_i$ with weights $s_i$
\begin{equation}
    \psi(\mathbf{r}) = \sum_i s_i h_i(\mathbf{r})\,.
\end{equation}
The stream function weights $s_i$ can be collected in a column vector $\mathbf{s} \in \mathbb{R}^{N_\text{v} \times 1}$. All operations in \mytexttt{bfieldtools} involving the stream function are linear, and can thus be represented as matrices operating on $\mathbf{s}$. For convenience, we will refer to $\mathbf{s}$ as the stream function from here on.

The surface-current density is obtained as the rotated gradient \citep{makinen_magnetic-field_2020} of the piecewise linear stream function, which makes it constant on each triangle face. Thus, we can express the current density $\mathbf{j} \in \mathbb{R}^{N_\text{f} \times 3}$ on the faces of the mesh as
\begin{equation}\label{eq:currentdensity_array}
    \mathbf{j}[i, j] = \sum_{k}^{N_\text{v}}\mathbf{G}^\perp[i, j, k] \mathbf{s}[k]\,,
\end{equation}
where $\mathbf{G}^\perp \in \mathbb{R}^{N_\text{f} \times 3 \times N_\text{v}}$ is the rotated gradient operator, which maps the scalar stream function defined on the $N_\text{v}$ mesh vertices to a 3-vector defined on the $N_\text{f}$ mesh faces. In Eq.~\ref{eq:currentdensity_array}, brackets are used to index individual elements of the operator. In practice, we represent the operators using multidimensional NumPy \mytexttt{ndarrays}, which are treated as a stack of 2D matrices, and with matrix multiplication applied with respect to their last two dimensions. Using \mytexttt{ndarrays}, Eq. \ref{eq:currentdensity_array} can be written in shorthand notation as
\begin{equation}\label{eq:currentdensity}
    \mathbf{j} = \mathbf{G}^\perp \mathbf{s}.
\end{equation}
In this paper, we use bold lower- and upper-case symbols, e.g. $\mathbf{r}$ and $\mathbf{R}$, to refer to column vectors and matrices, respectively.

In the stream-function representation of surface-current density, the magnetic field and other related quantities (such as the magnetic potentials) at given points are linear functions of the stream function. For example, knowing the coupling $\mathbf{C}_{\vec{B}} \in \mathbb{R}^{N_\text{p} \times 3 \times N_\text{v}}$ between the stream function values $\mathbf{s} \in \mathbb{R}^{N_\text{v} \times 1}$ defined at the $N_\text{v}$ vertices of the mesh and the magnetic field $\mathbf{B} \in \mathbb{R}^{N_\text{p} \times 3}$ at the $N_\text{p}$ field evaluation points $\mathbf{r} \in \mathbb{R}^{N_\text{p} \times 3}$, the magnetic field at $\mathbf{r}$ is computed as
\begin{equation}\label{eq:bfield}
\mathbf{B} = \mathbf{C}_{\vec{B}} \mathbf{s}\,.
\end{equation}

Quantities related to energy can be obtained with quadratic expressions of the stream function. Using the inductance matrix $\mathbf{M}$ [for definitions, see \citet{makinen_magnetic-field_2020}], the quadratic expression $\mathbf{s}^\top \mathbf{M} \mathbf{s}/2$ is the inductive field energy of the surface-current density. With the resistance matrix $\mathbf{R}$, the quadratic expression $\mathbf{s}^\top \mathbf{R} \mathbf{s}$ gives the Ohmic (heating) power of the surface current.

\subsection{Stream-function optimization}\label{sec:optim_intro}
\noindent When designing surface coils in the stream-function framework, one must find such an $\mathbf{s}$ that fulfills the given requirements. The problem can be formulated as an optimization task. A requirement for minimal energy or power can be convenient since the optimization problem then becomes convex and thus has an unique solution which can be solved efficiently. Other requirements for $\mathbf{s}$ can be included as inequality or equality constraints (e.g., one can constrain the magnetic field using Eq. \ref{eq:bfield}), thus maintaining the convexity when the constraint equations are linear. A solution can be found as long as the set of constraints defines a non-empty set of candidate solutions. Coil design is discussed in more detail in Section \ref{sec:coil_design}.

\subsection{Representations of fields and currents}\label{seq:bases_sph}
\begin{figure}[tb]
    \centering
    \includegraphics[width=\columnwidth]{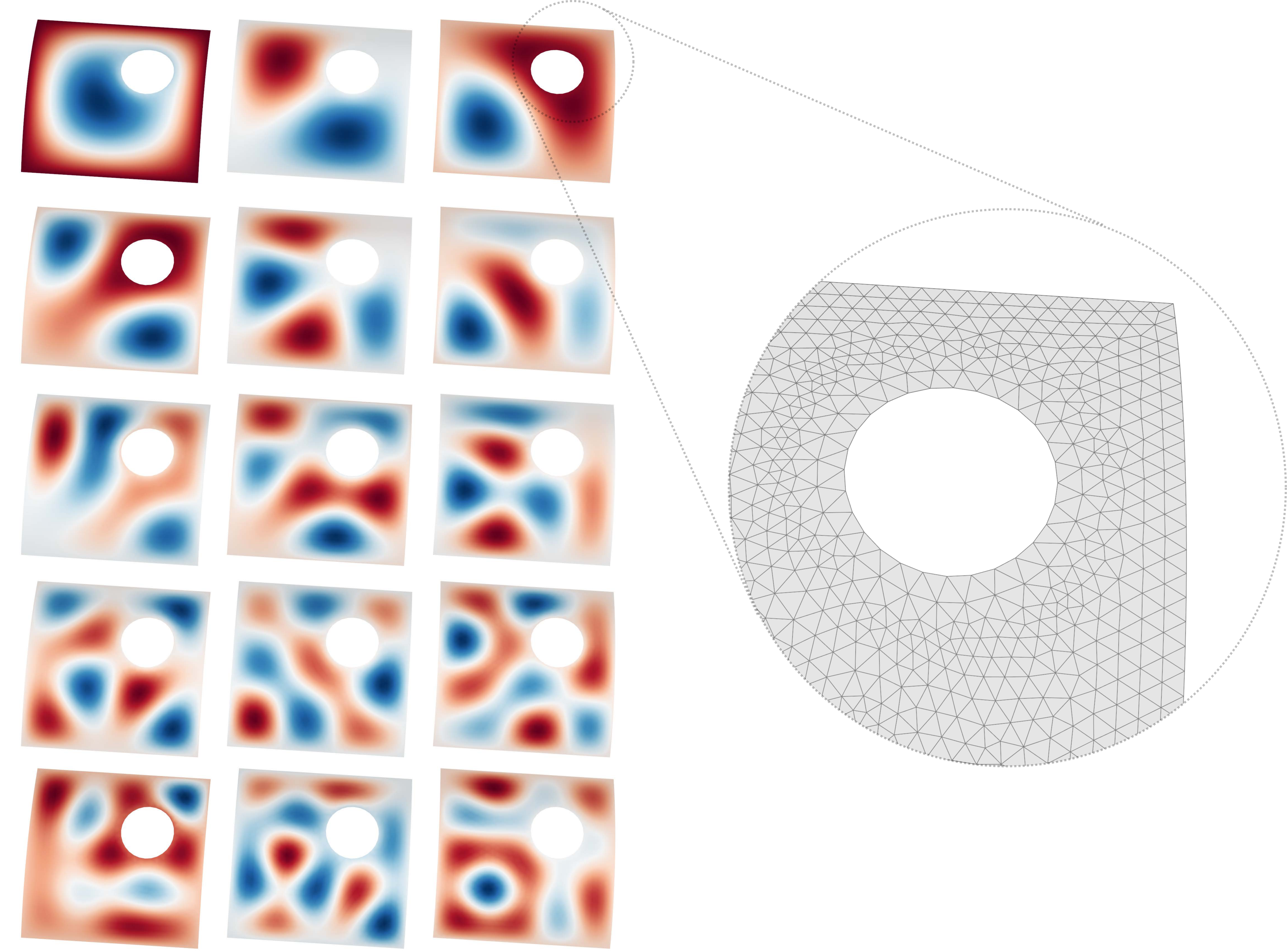}
    \caption{The first 15 surface harmonics of a triangle mesh representing a curved square with a hole. The index and thus spatial frequency increases from left to right, row by row. The tangential derivative is set to zero at the hole and outer boundaries. The mesh discretization is shown in the magnified inset on the right.}
    \label{fig:suh}
\end{figure}

\noindent In \mytexttt{bfieldtools}, the most flexible choice of basis for the stream function on a triangle mesh is arguably the direct use of the hat function basis, in which the surface current around each mesh vertex is described independently. Alternatively, one can apply the eigenfunctions of the surface Laplacian \citep{levy_laplace-beltrami_2006, reuter_discrete_2009}, which we call surface harmonics (SUH; Fig. \ref{fig:suh}; \citet{makinen_magnetic-field_2020}). The surface harmonics can be seen as a generalization of the more well-known spatial-frequency representations: on a sphere, the surface harmonics are essentially the spherical harmonics, and on a 2D plane, they correspond to the 2D Fourier series \citep{levy_laplace-beltrami_2006}. The series can represent smoothly-varying functions with a fairly small number of components, allowing the series to be truncated at a low order. For example, a stream function defined by the values on the 2000 vertices of a mesh might be expressed to a sufficient accuracy by 100 coefficients of the SUH series. Due to this compression, one can increase the mesh resolution without increasing the number of degrees of freedom and the computational cost, e.g., in optimization tasks. Truncating the SUH series also acts as an intuitive way to limit the maximum spatial frequency of the stream function and thereby in effect also its spatial gradient.

The surface harmonics are computed numerically using the generalized eigenvalue equation of the discretized surface-Laplacian operator $\mathbf{L}$ \citep{levy_laplace-beltrami_2006, reuter_discrete_2009}
\begin{equation}
    -\mathbf{L} \mathbf{V} = \mathbf{N} \mathbf{V} \mathbf{K}\,,
    \label{eq:suh_eigen}
\end{equation}
where $\mathbf{N}$ is a mass matrix taking into account the piecewise linear discretization of the mesh and $\mathbf{K} = \text{diag}(k_1^2 \dots k_{N_\text{h}}^2)$ contains the eigenvalues corresponding to the squared spatial frequencies of the $N_\text{h} \leq N_\text{v}$ surface harmonics, which are given by the columns of the basis matrix $\mathbf{V} \in \mathbb{R}^{N_\text{v} \times N_\text{h}}$. 

The SUH representation $\mathbf{a}$ of a stream function $\mathbf{s}$ can be obtained using the basis matrix $\mathbf{V}$ as $\mathbf{s} = \mathbf{V}\mathbf{a}$. Correspondingly, the magnetic field (Eq. \ref{eq:bfield}) can be computed directly from $\mathbf{a}$:
\begin{equation}\label{eq:suh_bfield}
\begin{array}{rl}
    \mathbf{B} = \mathbf{C}_{\vec{B}} \mathbf{s} = \mathbf{C}_{\vec{B}} \mathbf{V} \mathbf{a}\,.\\\end{array}
\end{equation}
Thus, the SUH coefficients $\mathbf{a}$ can be used to specify any field that can be produced by a surface current on the corresponding surface mesh.

Another way to represent the magnetic field is the spherical multipole series \citep{taulu_presentation_2005}. In this representation, the coefficients $\alpha_{lm}$ and $\beta_{lm}$ of the series can be used to specify the field in a source-free volume. The coefficients can be computed directly from the stream function with a linear mapping \citep{makinen_magnetic-field_2020}
\begin{equation}
\begin{array}{rl}
    \boldsymbol{\alpha} &= \mathbf{C}_\alpha\mathbf{s}\,, \\
    \boldsymbol{\beta} &= \mathbf{C}_\beta \mathbf{s}\,,
    \label{eq:multipoleS}
\end{array}
\end{equation}
where $\mathbf{C}_\alpha$ and $\mathbf{C}_\beta$ are the coupling of the stream function to the coefficient vectors $\boldsymbol{\alpha}$ and $\boldsymbol{\beta}$, respectively, containing the multipole coefficients indexed linearly with increasing $l$ and $m$ up to a predefined cutoff. As in Eq. \ref{eq:suh_bfield}, these coefficients can also be linearly mapped to the magnetic field as
\begin{equation}
\label{eq:multipoleB}
    \mathbf{B} = \mathbf{C}_{\vec{B}_\alpha}\boldsymbol{\alpha} + \mathbf{C}_{\vec{B}_\beta} \boldsymbol{\beta},
\end{equation}
where $\mathbf{C}_{\vec{B}_\alpha}$ and $\mathbf{C}_{\vec{B}_\beta}$ are the magnetic field coupling matrices representing spherical harmonic field components at the field evaluation points. This representation of the magnetic field is very compact for typical field profiles such as homogeneous or elementary gradient fields (which can be expressed with a single multipole coefficient), and can readily be applied, e.g., in coil design. To use the multipole series, the origin of the sphere used in the expansion has to be specified.

The SUH and multipole series can both provide a compact representation of the field. However, as discussed by Mäkinen and colleagues (\citeyear{makinen_magnetic-field_2020}), they have different convergence properties. The SUH and multipole coefficients can be fit to data, after which the estimated coefficients can be used to reconstruct and interpolate the magnetic field in the source-free space. In \mytexttt{bfieldtools}, we call the squared coefficients (both SUH and multipole) the spectrum of the magnetic field. 

\subsection{Boundary conditions}\label{sec:boundary_condition}
\noindent For the stream function to represent a divergence-free surface current (without current flowing in or out of the mesh), the derivative of the stream function along the boundaries of the mesh must be zero. In other words, the stream function must be constant on the boundary. It is typically convenient to set its value on the outer boundary of the mesh to zero. When the mesh has inner boundaries, the stream function value for the vertices on each boundary should be equal (but not necessarily zero). To enforce this, the hat functions along an inner boundary are combined into a single basis function, the value of which is constant along the boundary.

\subsection{Eddy currents}\label{seq:eddy}
\noindent There are many ways to control eddy-current-induced fields in a region of interest when quickly switching the applied magnetic field \citep[e.g.][]{peeren_stream_2003, zevenhoven_conductive_2014, zevenhoven_dynamical_2015}. Here, we present a way to compute the secondary field caused by eddy currents induced in some nearby conductor due to a primary field generated by a current in a surface coil. For an idealized case where the current waveform is a (Heaviside) step function, the instantaneous induced magnetic field $\mathbf{\vec B}_2$ caused by the eddy currents within a region of interest at time point $t$ is \citep{peeren_stream_2003-1}
\begin{equation}\label{eq:eddy}
    \mathbf{B}_2(t) = - \mathbf{C}_{\vec{B}_2} \mathbf{U} e^{-\mathbf{D} t} \mathbf{U}^{-1} ( \mathbf{M}_{12} \mathbf{M}_{2}^{-1})^\top \mathbf{s}_1\,,
\end{equation}
where $\mathbf{C}_{\vec{B}_2}$ is the magnetic field coupling matrix of the conductive object to the region of interest, $\mathbf{M}_{12}$ is the mutual inductance matrix between the coil mesh and the conductor mesh, and $\mathbf{M}_{2}$ is the self-inductance matrix of the conductor mesh. Matrices $\mathbf{U}$ and $\mathbf{D}=\text{diag}(1/\tau_1, \dots, 1/\tau_N) $ are determined by the generalized eigenvalue problem
\begin{equation}
\mathbf{R}_2 \mathbf{U} = \mathbf{M}_2 \mathbf{U} \mathbf{D}\,,
\end{equation}
where $\mathbf{R}_2$ is the resistance matrix of the conductor and $\tau_1, \dots, \tau_N$ are the time constants of the $N$ different eddy current modes corresponding to the columns of $\mathbf{U}$. 

\subsection{Magnetic shielding}\label{seq:shielding}
\noindent High-permeability shields are often used to minimize the effect of ambient magnetic fields on sensitive systems or experiments. However, the shield also distorts any magnetic fields generated inside the shield. When the relative permeability of the shield is high, the effect of the shield can be approximated by the boundary condition that the magnetic scalar potential on the inner shield surface is constant \citep{makinen_magnetic-field_2020}. We solve this boundary condition by setting the constant to zero and by introducing an equivalent stream function $\mathbf{s}_2$ to the shield surface such that
\begin{equation}
    \mathbf{C}_{U_1}\mathbf{s}_1 = -\mathbf{C}_{U_2}\mathbf{s}_2,
\end{equation}
where $\mathbf{C}_{U_1}$ and $\mathbf{C}_{U_2}$ are the magnetic scalar potential coupling matrices of the coil and the shield for collocation points slightly inside the mesh. With the equipotential boundary condition, the magnetic field expression takes the form
\begin{equation}\label{eq:shielding}
    \mathbf{B} = \mathbf{C}_{\vec{B}_1}\mathbf{s}_1 + \mathbf{C}_{\vec{B}_2}\mathbf{s}_2 = (\mathbf{C}_{\vec{B}_1} -  \mathbf{C}_{\vec{B}_2} \mathbf{C}_{U_2}^{-1}  \mathbf{C}_{U_1})\mathbf{s}_1\,,
\end{equation}
which allows for the effect of magnetic shielding to be computed for any surface-current density within the shield.

\section{Software overview}
\noindent \mytexttt{bfieldtools} is implemented purely in Python and leverages a large number of packages and libraries within the open-source scientific Python community. We use the trimesh package \citep{dawson-haggerty_michael_trimesh_2020} for all mesh-related functionality. For numerical operations and linear algebra, we use NumPy \citep{oliphant_guide_2015} and SciPy \citep{oliphant_python_2007, virtanen_scipy_2020}. Visualizations are generated using matplotlib \citep{hunter_matplotlib:_2007} and mayavi \citep{ramachandran_mayavi:_2011} in two and three dimensions, respectively. The quadpy package \citep{schlomer_nschloequadpy_2020} is used for quadrature scheme generation for numerical integration, and the CVXPY package \citep{diamond_cvxpy:_2016, agrawal_rewriting_2018} is employed in coil optimization.

\mytexttt{bfieldtools} has extensive online documentation, generated using Sphinx (\url{https://www.sphinx-doc.org}). The documentation includes an API reference, a large number of application examples acting as tutorials as well as links to background literature. 

\subsection{Software components}
\begin{figure}
    \centering
    \includegraphics[width=\columnwidth]{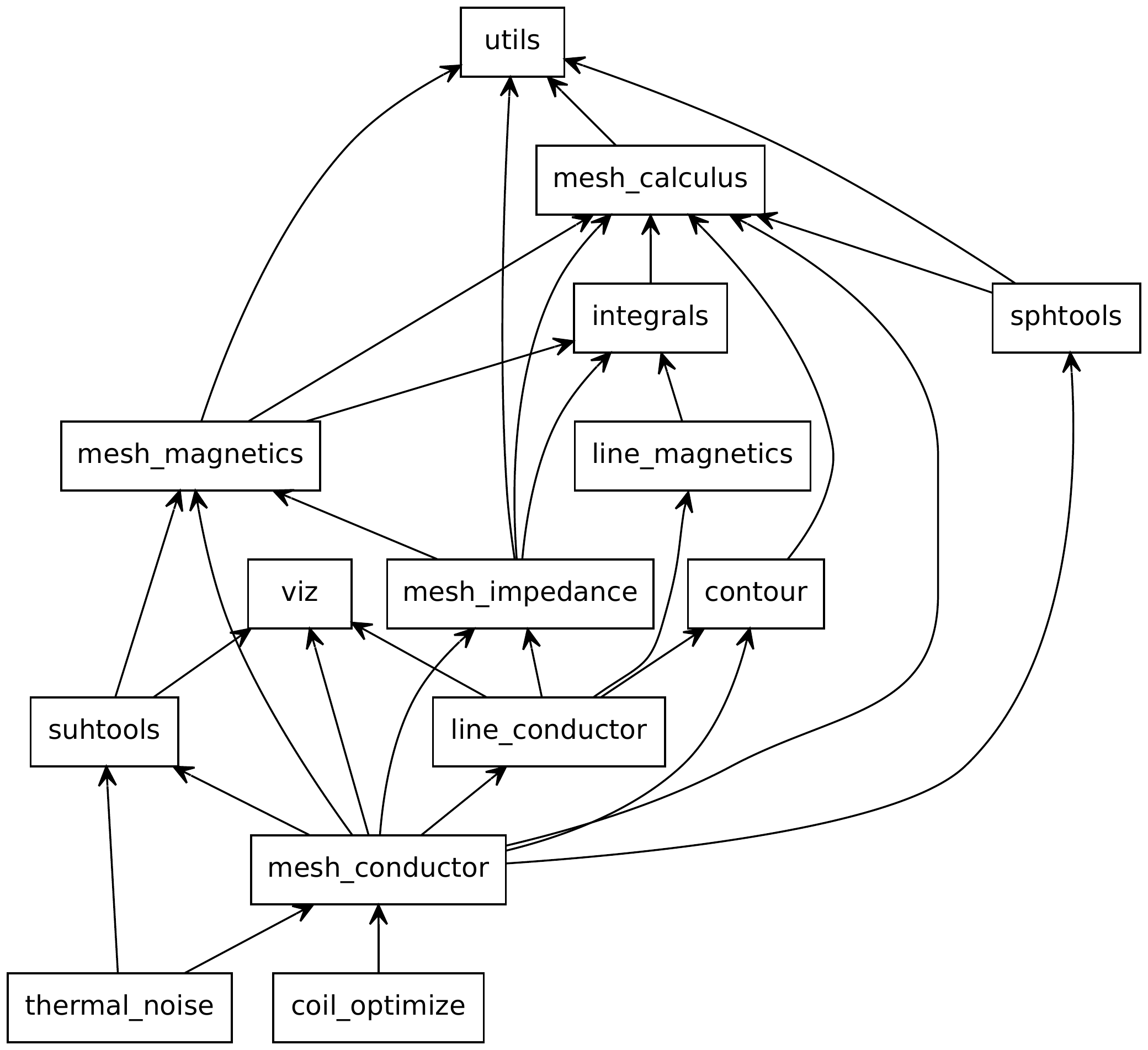}
    \caption{Submodule relations of \mytexttt{bfieldtools} software.}
    \label{fig:UML}
\end{figure}

\noindent In this section, we summarize the functionality of the individual software submodules in \mytexttt{bfieldtools}. A graphical overview of the relations of the submodules in the package and the general software architecture is shown in Fig. \ref{fig:UML}.

\paragraph{Mesh conductor class}
A large portion of the user-facing functionality of \mytexttt{bfieldtools} is centered around the use of \mytexttt{MeshConductor} objects that encapsulate a Trimesh triangle mesh object. The \mytexttt{MeshConductor} wrapper adds properties and methods such as the surface stream function, inductance and resistance matrices as well as magnetic field and potential coupling matrices. These properties are implemented with lazy evaluation, i.e. they are computed only when needed. In addition, the computed coupling matrices are cached to minimize redundant computation. Since the stream function may be represented in several different bases, the \mytexttt{MeshConductor} object is implemented such that it can internally handle basis changes.

\paragraph{Integrals}
This submodule forms the core of \mytexttt{bfieldtools}. Using the analytical integral equations implemented in this submodule, most physical quantities used in the software package can be computed without numerical quadratures, yielding better accuracy in the vicinity of the mesh. In a typical use case, these functions are not directly accessed. Instead, they are used as building blocks for the mesh operators in other submodules. For an in-depth description of the analytical integrals, see \citet{makinen_magnetic-field_2020}.

\paragraph{Mesh calculus}
This submodule implements calculus functions for triangle meshes. These functions include the gradient and rotated gradient, which map a scalar field represented as vertex values to tangential vector fields represented as 3-vectors on the mesh faces. The submodule also contains surface divergence and curl functions, which map tangential vector values on the faces to scalar values on the vertices \citep[for more details see e.g.][]{botsch_polygon_2010, de_goes_vector_2016, reusken2018stream}. Based on the same framework, the surface-Laplacian (or Laplace--Beltrami) operator $\mathbf{L}$ (acting on scalars and vector functions defined on the vertices) is also implemented. Due to the boundary conditions discussed in Section \ref{sec:boundary_condition}, the dimensions of the Laplacian differ for closed and open surfaWe present a novel open-source Python software package, bfieldtools, for magneto-quasistatic calculations with current densities on surfaces of arbitrary shape. The core functionality of the software relies on a stream-function representation of surface-current density and its discretization on a triangle mesh. Although this stream-function technique is well-known, to date the related software implementations have not been published or have been limited to specific applications. With bfieldtools, we aimed to produce a general, easy-to-use and well-documented open-source software. The software package is written purely in Python; instead of explicitly using lower-level languages, we address computational bottlenecks through extensive vectorization and use of the NumPy library. The package enables easy deployment, rapid code development and facilitates application of the software to practical problems. In this paper, we describe the software package and give an extensive demonstration of its use with an emphasis on one of its main applications – coil design.ces as well as for surfaces with holes, where each hole corresponds to one free value in the system. The mesh calculus module also includes a function to compute the mesh mass matrix $\mathbf{N}$ used, e.g., in Eq.~\ref{eq:suh_eigen}.

\paragraph{Mesh magnetics}
This submodule contains functions for computing the coupling matrices of the magnetic field, vector potential and scalar potential ($\mathbf{C}_{\vec B}$, $\mathbf{C}_{\vec A}$ and $\mathbf{C}_U$, respectively) described in Section \ref{sec:stream}. In order to trade time for memory usage, the functions in this submodule include an option to compute the matrices in serial chunks. Furthermore, one may compute the magnetic field either using analytic integrals or, to speed up the computation, using numerical quadratures. When using the quadrature implementation, the choice of quadrature scheme can be freely specified by the user.

\paragraph{Mesh impedance}
This submodule includes computations of, e.g., the resistance matrix $\mathbf{R}$, the (self-)inductance matrix $\mathbf{M}$, the mutual inductance matrix $\mathbf{M}_{12}$ between two meshes as well as the mutual inductance between a mesh and loops of line currents (represented with connected current segments; see the \emph{line magnetics} submodule). The inductance matrices are computed using the magnetic vector potential, as implemented in the mesh magnetics submodule. As these functions are highly vectorized, they are fast but require a significant amount of memory. As with the coupling matrix functions, the inductance matrix computations can be computed in serial chunks to save memory. Furthermore, one may speed up the computation by applying a numerical quadrature for evaluation points further away than a user-specified number of average triangle side lengths.

\paragraph{Contouring}
For surface coil design, the contouring submodule contains functions for extracting discrete current loops from the continuous surface current density. The submodule also contains functions to process and smooth these current loops.

\paragraph{Line magnetics}
\mytexttt{bfieldtools} includes a module for computations related to polyline currents comprising connected line segments. This includes the generated magnetic field as well as magnetic vector and scalar potentials. In addition, the module provides functions for computing (mutual) inductance of current loops.

\paragraph{Visualization}
This submodule contains a variety of functions for visualizing meshes, stream functions, current loops, as well as fields and potentials. These functions are mainly wrappers for matplotlib (2D) and mayavi (3D) with suitable defaults for the type of data being plotted.

\paragraph{Spherical harmonics}
\mytexttt{bfieldtools} includes a submodule containing functions for generating real (vector) spherical harmonics of arbitrary order as well as functions for multipole representation of the magnetic field (Eq. \ref{eq:multipoleB}). Additionally, functions to, e.g., visualize the spherical harmonics, to estimate the spherical harmonics coefficients from data and to compute the multipole coefficients from the mesh stream function (Eq. \ref{eq:multipoleS}) are provided. Different normalization schemes for the coefficients are provided.

\paragraph{Surface harmonics}
The surface harmonics submodule contains tools for generating SUH components on a surface as well as working with SUH function expansions. These features are implemented in a class that calculates the expansion truncated to a given number of components. The returned object can be used for calculating the magnetic field associated with the SUH basis functions, estimating the SUH coefficients from data, as well for visualizing the functions.

\paragraph{Thermal noise}
\mytexttt{bfieldtools} also includes a module for computing thermal AC magnetic noise arising from thin conducting objects (modeled using triangle meshes). The noise calculation implemented in the submodule uses the same computational stream-function framework and is described in more detail by \citet{iivanainen_general_2020}.

\paragraph{Coil optimization}
This submodule provides wrapper functions for quadratic coil optimization either using the regularized least-squares method or a numerical iterative solver via CVXPY. The functions take easy, human-readable parameters for the coil specification and constraints.

\paragraph{Utilities}
Finally, \mytexttt{bfieldtools} includes a separate submodule for a variety of helper functions and utilities that are used across the other submodules.

\section{Coil design}\label{sec:coil_design}
\noindent One of the main applications of the \mytexttt{bieldtools} software package is coil design. There are many applications in which one needs to design a coil fulfilling a set of requirements, e.g., on the field profile or homogeneity, the mechanical dimensions of the coil, stray field, coil heating and manufacturability. As discussed in Section \ref{sec:optim_intro}, the coils can be designed by optimizing a stream function such that a quadratic expression is minimized while some additional linear constraint(s) are met.

\subsection{Optimization methods}
\noindent Depending on how the optimization problem is formulated, different optimization methods can be applied. In \mytexttt{bfieldtools}, the main optimization method for coil design is constrained quadratic optimization using a numerical iterative solver. The use of a numerical solvers allows the use of linear inequality constraints (such as allowing for a specific tolerance in, e.g., field profile). In \mytexttt{bfieldtools}, we employ the CVXPY convex optimization modeling language for accessible and easily applied optimization. When using CVXPY for optimizing the stream function $\mathbf{s}$, the problem statement is of the form
\begin{equation}\label{eq:qp}
\begin{array}{rl}
\mathrm{minimize} & \frac{1}{2} \mathbf{s}^\top \mathbf{P} \mathbf{s} + \mathbf{q}^\top \mathbf{s} \\
\mathrm{subject\ to} & \mathbf{G} \mathbf{s} \leq \mathbf{h}, \\
    & \mathbf{A} \mathbf{s} = \mathbf{b}\,, \\
\end{array}
\end{equation}
where $\mathbf{P}$ is the quadratic objective matrix (e.g.,  inductance $\mathbf{M}$ or resistance $\mathbf{R}$), $\mathbf{q}$ defines an optional linear penalty term, and the linear equality and inequality constraints are applied as needed. Multiple simultaneous constraints of the same type can easily be applied by stacking the constraint matrices. Furthermore, due to the flexibility of the CVXPY framework, one may also include additional constraints, such as constraining the $p$-norm [e.g. 1-norm or $\infty$-norm, as done by \citet{poole_convex_2014}] of a linear expression for $\mathbf{s}$ $\Vert \mathbf{D} \mathbf{s} - \mathbf{t}\Vert_p \leq e$, or by constraining the stream function value of specific vertices to be equal: $ \mathbf{s}_i = \mathbf{s}_j$.
    
An alternative approach \citep[e.g.][]{pissanetzky_minimum_1992} is to formulate the problem as a quadratic optimization without hard constraints, and instead use trade-off parameters. In this form, an example problem is formulated as 
\begin{equation}
\begin{array}{rl}
\mathrm{minimize} & \frac{1}{2} \mathbf{s}^\top \mathbf{P} \mathbf{s} + \lambda\|\mathbf{q}-\mathbf{C}_{q} \mathbf{s}\|^2\,,
\end{array}
\end{equation}
where $\mathbf{q}$ determines the desired values of some quantity in some number of points and $\mathbf{C}_{q}$ is the coupling matrix for that quantity and those points. Finally, $\lambda$ is a scalar trade-off parameter, weighting the solution either towards minimizing the primary objective function or a penalty function. This formulation has a closed-form solution
\begin{equation}\label{eq:tikhonov-lsq}
\begin{split}
    \mathbf{s} &= \left(\mathbf{C}_{q}^\top \mathbf{C}_{q} + \frac{1}{\lambda} \mathbf{P}\right) ^{-1} \mathbf{C}_{q} \mathbf{q}\\
    &= \mathbf{P}^{-1} \mathbf{C}_{q}^\top \left(\mathbf{C}_{q} \mathbf{P}^{-1} \mathbf{C}_{q}^\top + \lambda \mathbf{I}\right)^{-1}\mathbf{q}\,,
\end{split}
\end{equation}
which may be familiar as the Tikhonov-regularized least-squares formula. In the general case, multiple quadratic penalty terms may be applied, each with their own $\lambda_i$. In order to include linear equality constraints, one may, e.g., employ a Lagrange multiplier method as done by \citet{lemdiasov_stream_2005} and \citet{poole_improved_2007}. However, while having good performance in problems with straightforward constraints, this inversion-based approach cannot accommodate hard inequality constraints.

\subsection{Objective functions}\label{seq:objs}
\noindent In \mytexttt{bfieldtools}, two main options for the quadratic objective are directly implemented. These are the minimization of the resistive power or the magnetic energy. Minimizing the resistive power $\mathbf{s}^\top\mathbf{R}\mathbf{s}$ results in a maximally smoothly varying stream function, as well as minimizing the resistive losses in the coil. This reduces the need for cooling the coil when large currents are used. Minimizing the magnetic energy $\mathbf{s}^\top\mathbf{M}\mathbf{s}/2$ results in minimal inductance of the coil. This reduces the voltage involved in fast ramping of the current in the coil.

Functions for magnetic and resistive energy minimization typically result in fairly similar stream functions. The two functions differ in that magnetic energy minimization allows for somewhat more variation at higher spatial frequencies of the stream function. These would be penalized more in resistive energy minimization. One may also form the quadratic objective as a weighted combination of resistive power and magnetic energy. Finally, \mytexttt{bfieldtools} allows for use of any other user-specified quadratic objective function.

\subsection{Constraints}\label{seq:constraints}
\noindent Minimizing the quadratic objective without any penalty terms or constraints would lead to a trivial zero-current, zero-field solution. Thus, one must specify additional constraints to determine the final current pattern.

In coil design, constraining the magnetic field within a target region is the most typical constraint. In addition to specifying a target field one may also want to explicitly limit the stray field outside the coil. Using the spherical harmonics representation of the magnetic field as presented in Section \ref{seq:bases_sph}, one can also place constraints on $\boldsymbol{\alpha}$ and $\boldsymbol{\beta}$. Using a multipole-based constraint for the magnetic field naturally satisfies Maxwell's equations in a source-free volume, whereas multiple user-specified point-based field constraints are not guaranteed to do so. One may also add other constraints, e.g.\ related to eddy currents (see Section \ref{seq:eddy}) or to high-permeability shielding (see Section \ref{seq:shielding}).

The use of inequality constraints in the optimization, as is possible when using an iterative solver, allows directly specifying the desired properties of the coil. This may be more intuitive than the use of trade-off/penalty parameters employed in the least-squares formulation. The use of inequality constraints also allows for wiggle room in the coil specification. This wiggle room decreases the need for apodization \citep[e.g.][]{forbes_novel_2004, hidalgotobon_theory_2010}, i.e.\ post-optimization smoothing of the stream function. Apodization has been applied due to high spatial-frequency oscillations or 'ringing' in the stream function, which may arise when a target-field equality constraint is used, especially when minimizing the magnetic energy. 

More sophisticated methods to limit high-frequency ringing directly constrain the gradient of the stream function; the spatial gradient of the stream function defines the surface-current density, and by extension, the spacing of the discretized coil windings. Constraining the maximum gradient affects the minimum spacing of windings, which can also be useful with regards to manufacturability. Limiting the maximum current density can also decrease local heating issues in high-power applications. The minimax $|j|$ method presented by \citet{poole_minimax_2010, poole_minimax_2012} should be similar in effect to constraining the stream function gradient, but works somewhat differently from an optimization viewpoint. An alternative way to reduce the minimum spacing of windings is to use a truncated SUH basis limited to low spatial frequencies.

\subsection{Discretization to wire segments}
\noindent The surface-current density is obtained from the optimized stream function with Eq.~\ref{eq:currentdensity}. To extract the geometry of discrete conductor loops, one can simply use any number of stream function isolines with equal spacing in terms of stream function value. The number of isolines, i.e. current loops, can be freely chosen to fit the application; more loops will naturally result in a larger magnetic field per unit current and larger inductance, but will also better approximate the continuous surface current, thus having a smaller discretization error. Finally, the independent current loops must be connected in series, with special care taken to ensure that the current direction corresponds to the continuous current density. The manner in which the loops are connected should depend on manufacturing method and scale. For example, on a printed circuit board, the loops may be connected using vias and multiple layers, while larger-scale coils may even use soldered wire segments.

\section{Examples}
\noindent The online documentation of \mytexttt{bfieldtools} (available at \url{https://bfieldtools.github.io}) contains several examples of applications, with code and accompanying explanatory text and figures. In this section, we discuss a number of selected examples in detail, walking through some of the software workflow, design decisions and rationale. However, for brevity and to focus on the essentials, we omit most imports as well as some repetitive or trivial steps. Online examples will be provided in full.

\subsection{Biplanar coils with minimal stray field}
\noindent In this example, we design a biplanar coil which produces homogeneous field within a spherical target region between the two square surface coils. In addition, we explicitly minimize the stray field on a spherical surface surrounding the coils. We start by importing the mesh file into a \mytexttt{MeshConductor} object. In this example, we use a very dense mesh, with 12 442 vertices and 24 304 faces. To speed up computation and limit the coil winding density, we use a truncated SUH representation for the stream function with $N_\text{h}=100$.

\begin{lstlisting}[language=Python]
coil = MeshConductor(mesh_file=path, 
                 basis_name='suh',
                 N_suh=100)
\end{lstlisting}

\begin{figure*}[htb]
    \centering
    \includegraphics[width=\linewidth]{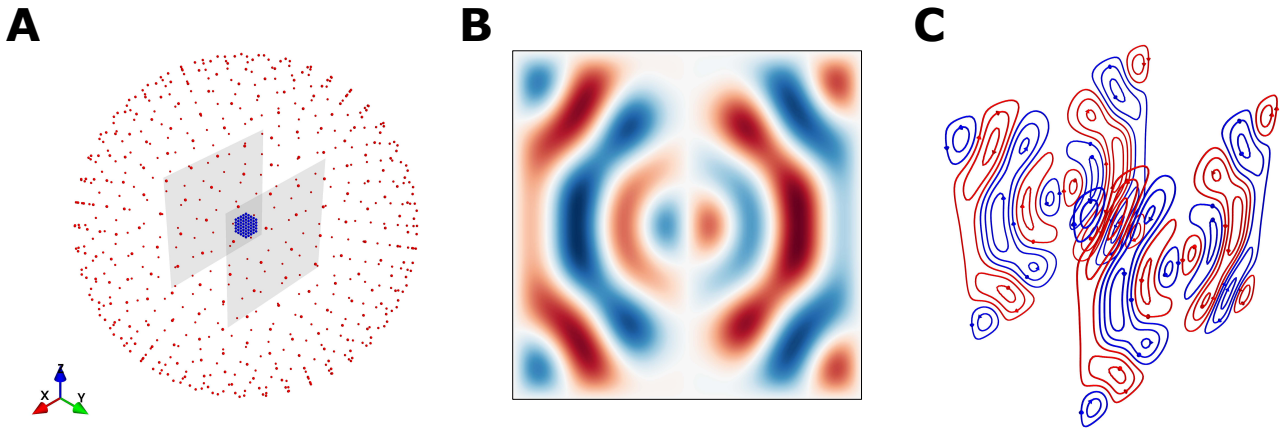}
    \caption{A: Biplanar coil surface meshes, target points (in blue) and stray field points (in red). B: Optimized stream function on one of the coil planes generating homogeneous magnetic field along the X-axis. C: Discretized coil windings.}
    \label{fig:biplanar}
\end{figure*}

We omit code lines for the generation of target and stray field points, and instead visualize the whole geometry in Fig. \ref{fig:biplanar}A. The target points are on a grid within a sphere around the centre of the biplanar coil (diameter 0.3 times the square side length), and the stray field points are on a spherical surface surrounding the coils (radius twice the square side length). 

After having generated the geometry, we set the field specification at the defined target and stray points. In this case, we specify a homogeneous field along the $x$-axis (within the target volume). We allow for <0.5\% deviation in all three Cartesian components of the field at all target points. For all components of the stray field, we allow for <1\% deviation from the target field amplitude. The homogeneous target field amplitude is set to a numerical value of $1$ for convenience. The absolute value does not matter, and will be scaled appropriately in the numerical solver. Having set the field specifications, we can run the numerical solver to optimize the stream function. We use the Ohmic power as the primary quadratic penalty and apply linear inequality constraints on the magnetic field at the target and stray field points.

\begin{lstlisting}[language=Python]
import numpy as np  # Shorthand for numpy

target_field = np.zeros_like(target_pts)
target_field[:, 0] += 1 # Homogeneous field along x-axis

target_spec = {'coupling': coil.B_coupling(target_pts),
               'abs_error': 0.005, # 0.5% error
               'target': target_field}

stray_spec = {'coupling': coil.B_coupling(stray_pts),
              'abs_error': 0.01, # 1% error
              'target': np.zeros_like(stray_pts)}

s, problem = optimize_streamfunctions(coil, [target_spec, stray_spec], objective='minimum_ohmic_power')
\end{lstlisting}
Having computed the optimized stream function (see Fig.~\ref{fig:biplanar}B), we can now convert the continuous stream function (current density) into discrete current loops and plot the result (as seen in Fig.~\ref{fig:biplanar}C). 

\begin{lstlisting}
loops, loop_values = scalar_contour(coil.mesh, 
                                    coil.s,
                                    N_contours=6)
                                    
plot_3d_current_loops(loops, colors='auto', figure=f)
\end{lstlisting}

\subsection{Eddy current minimization}
\noindent Here, we use a geometry with a cylindrical coil surface similar to a conventional MRI bore, surrounded by a larger conducting cylindrical RF shield (both cylinder meshes have 4 764 vertices and 9 368 faces). We will design a reference coil which generates a homogeneous field along the X-axis (perpendicular to the long axis of the cylinder) within a spherical target volume. Furthermore, we will compute the eddy currents produced in the RF shield when switching the current in the coil. We also design another coil whose excitation generates minimal eddy-current field transients in the target volume. To this end, we add the expression in Eq.~\ref{eq:eddy} as a constraint to the coil optimization procedure. We specifically do not use a an outer set of coils for self-shielding in order to showcase the eddy-current-induced field minimization procedure.

\begin{figure*}[htb]
    \centering
    \includegraphics[width=\linewidth]{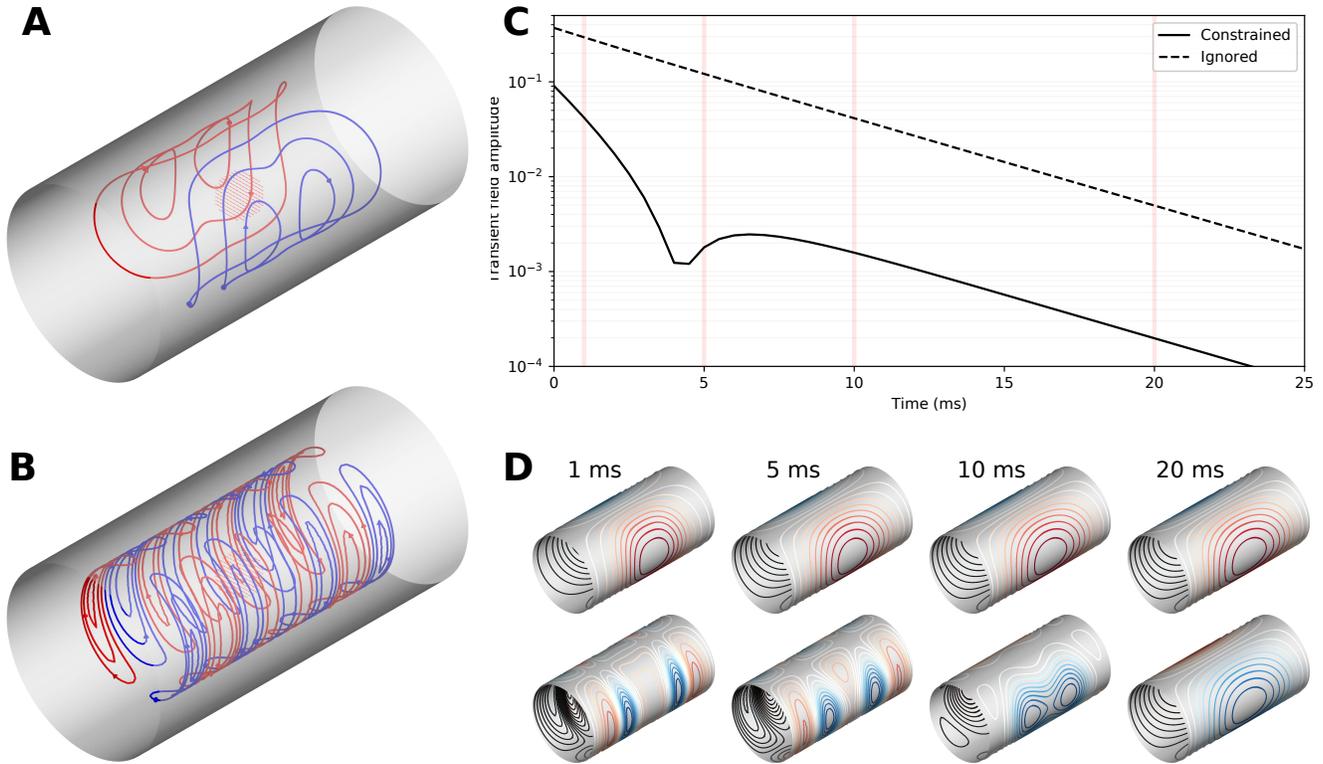}
    \caption{Cylindrical coils that are designed (A) ignoring or (B) minimizing the transient field that is generated by eddy currents in a surrounding cylindrical conductor when switching the current in the coil. Loop color and arrows indicate the direction of the current. (C) Mean transient field amplitude induced into the region of interest. Amplitudes are expressed as a fraction of the homogeneous field generated by the coil. (D) Eddy-current patterns induced into the cylindrical conductor at the time points indicated by vertical lines in C. The upper row corresponds to the coil in A, and the lower row to that in B.}
    \label{fig:eddy_currents}
\end{figure*}

We omit the preparation steps shown in the previous example and instead present the geometry in Fig.~\ref{fig:eddy_currents}. First, we compute the eddy-current modes and time constants of the cylindrical shield. As no current enters or leaves the shield, we set the stream function to zero at the boundary. The mesh boundary vertices are then not included in the generalized eigenvalue problem of Eq.~\ref{fig:eddy_currents} and the entries in $\mathbf{U}$ corresponding to boundary vertices are fixed to zero by setting the \mytexttt{MeshConductor} object basis to 'inner' (meaning inner vertices only). In this example, we only compute the 500 longest-lived eddy current modes, as the faster modes will have negligible effects past 1 ms.
\begin{lstlisting}[language=Python]
from scipy.linalg import eigh

#Values for 0.5 mm thick aluminium at room temperature
shield = MeshConductor(mesh_file=path, basis_name='inner',
                   resistivity=2.8e-8, #Ohm*meter
                   thickness=0.5e-3 #meter) 

#Compute 500 longest-lived eddy current modes
l, U = eigh(shield.resistance, 
            shield.inductance,
            eigvals=(0, 500))
\end{lstlisting}

Knowing the eddy-current dynamics, we can now define the coil-design specification and run the optimization procedure. In the static case, we allow for <0.5\% field deviation on all field components from the target field at the target points. Additionally, we limit all components of the eddy\hyp{}current\hyp{}induced transient field at the target points at time points 1 ms, 3 ms and 5 ms to <5\%, <1\% and <0.25\% of the homogeneous field strength, respectively. 

\begin{lstlisting}[language=Python]
M_coupling = np.linalg.solve(-shield.inductance, mutual_inductance.T) #Inductance part of Eq. 9

time = [0.001, 0.003, 0.005] #Time points in seconds
abs_error = [0.05, 0.01, 0.0025] #Error limits

#Initialize specification list
induction_spec = []

for idx, t in enumerate(time):
    time_decay = U @ np.diag(np.exp(-l*t)) @ np.linalg.pinv(U) #Time decay part of Eq. 9
    
    eddy_coupling = shield.B_coupling(target_pts) @ time_decay @ M_coupling #Eq. 9 put together
    
    zeros = np.zeros_like(target_field)
    induction_spec.append({'coupling':eddy_coupling,
                           'abs_error':abs_error[idx],
                           'target':zeros})

target_spec = {'coupling':coil.B_coupling(target_pts),
               'abs_error':0.005, #0.5% error 
               'target':target_field}

coil.s, coil.problem = optimize_streamfunctions(coil, [target_spec] + induction_spec, objective='minimum_inductive_energy')
\end{lstlisting}

First, for the reference coil, we omit the eddy\hyp{}current\hyp{}related parts of the coil specification. Then, for the second coil, we include the eddy-current constraints. The resulting discretized windings for the two coils are shown in Fig.~\ref{fig:eddy_currents}A\&B. In Fig.~\ref{fig:eddy_currents}C, it is evident how much the eddy currents are suppressed for the second coil. The eddy-current field decays to below 1\% of the static target field in 2.6 ms, whereas for the reference coil this time is 16.7 ms. Furthermore, Fig.~\ref{fig:eddy_currents}D also shows the eddy-current pattern on the shield surface at different time points. For the coil in Fig.~\ref{fig:eddy_currents}B, the eddy currents initially take such a pattern that they do not induce field in the target region. However, over time, the eddy-current pattern spreads and eventually resembles that of the reference coil in Fig.~\ref{fig:eddy_currents}A, as the longer-lived eddy-current modes also have lower spatial frequencies.

\subsection{Interactions with magnetic shielding}
\begin{figure*}[htb]
    \centering
    \includegraphics[width=0.8\linewidth]{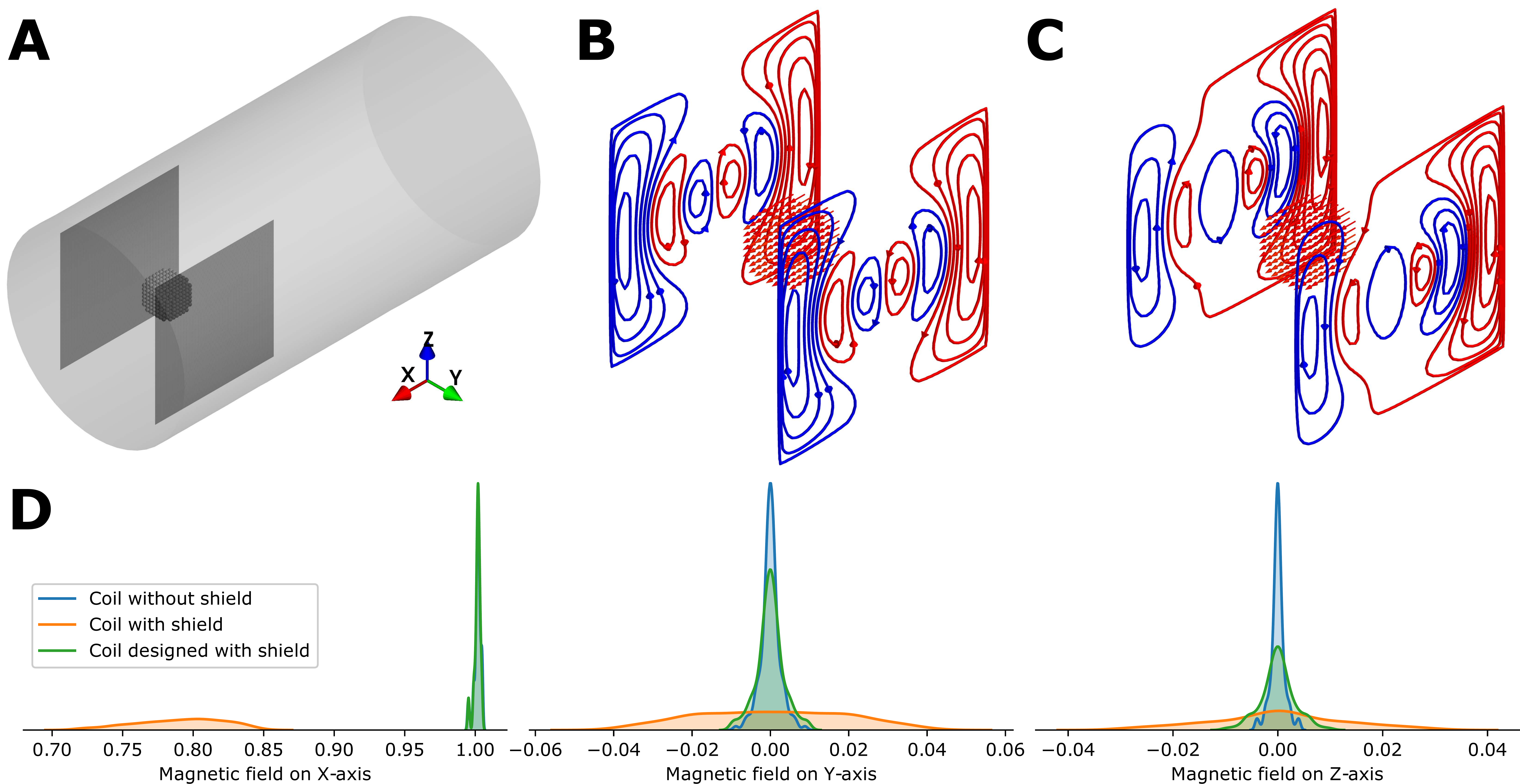}
    \caption{A: Biplanar coil within an ideal cylindrical ideal magnetic shield. B: Coil designed to produce a homogeneous magnetic field along the axis of the cylinder, but with the effect of the shield neglected. C: Coil designed while prospectively taking the effect of the shield into account. D: Component-wise field distributions of the coils at the target points with and without taking the shield into account.}
    \label{fig:shielded}
\end{figure*}

\noindent We consider the same square coil surfaces as in the first example, except now within a closed cylindrical magnetic shield (2 773 mesh vertices and 5 542 faces), see Fig.~\ref{fig:shielded}A. To emphasize the field distortion caused by the shield, we place the coils very close to the cylinder end. Again, we omit the preparations and only present the steps that lead to a coil in which the effect of the magnetic shielding is prospectively taken into account. We start by defining collocation points slightly inside the shield surface. We continue by solving Eq.~\ref{eq:shielding}, corresponding to the equipotential boundary condition at the shield (or at the collocation points). We include the solved field distortion in the coupling in the target field coupling matrix. In the optimization, we apply a linear inequality constraint for the target field.

\begin{lstlisting}[language=Python]
# Collocation points slightly inside the shield
shield_points = shield.mesh.vertices - epsilon*shield.mesh.vertex_normals

# Shielding distortion, as in Eq. 12
shield_B_distortion = shield.B_coupling(target_pts) @ np.linalg.solve(shield.U_coupling(shield_points), coil.U_coupling(shield_points))

total_coupling = coil.B_coupling(target_pts) + shield_B_distortion

target_spec_w_shield = {'coupling': total_coupling,
                        'abs_error': 0.01,
                        'target': target_field}

coil.s, coil.problem = optimize_streamfunctions(coil, [target_spec_w_shield], objective='minimum_inductive_energy')
\end{lstlisting}

The resulting coil windings are shown in Fig.~\ref{fig:shielded}C together with a reference coil design, for which the effect of the high-permeability shield was neglected (Fig.~\ref{fig:shielded}B). The field distribution within the target region is shown in Fig.~\ref{fig:shielded}D.

\subsection{Magnetic field interpolation using equivalent surface currents}\label{sec:suh_interp}
\begin{figure}[htb]
    \centering
    \includegraphics[width=\linewidth]{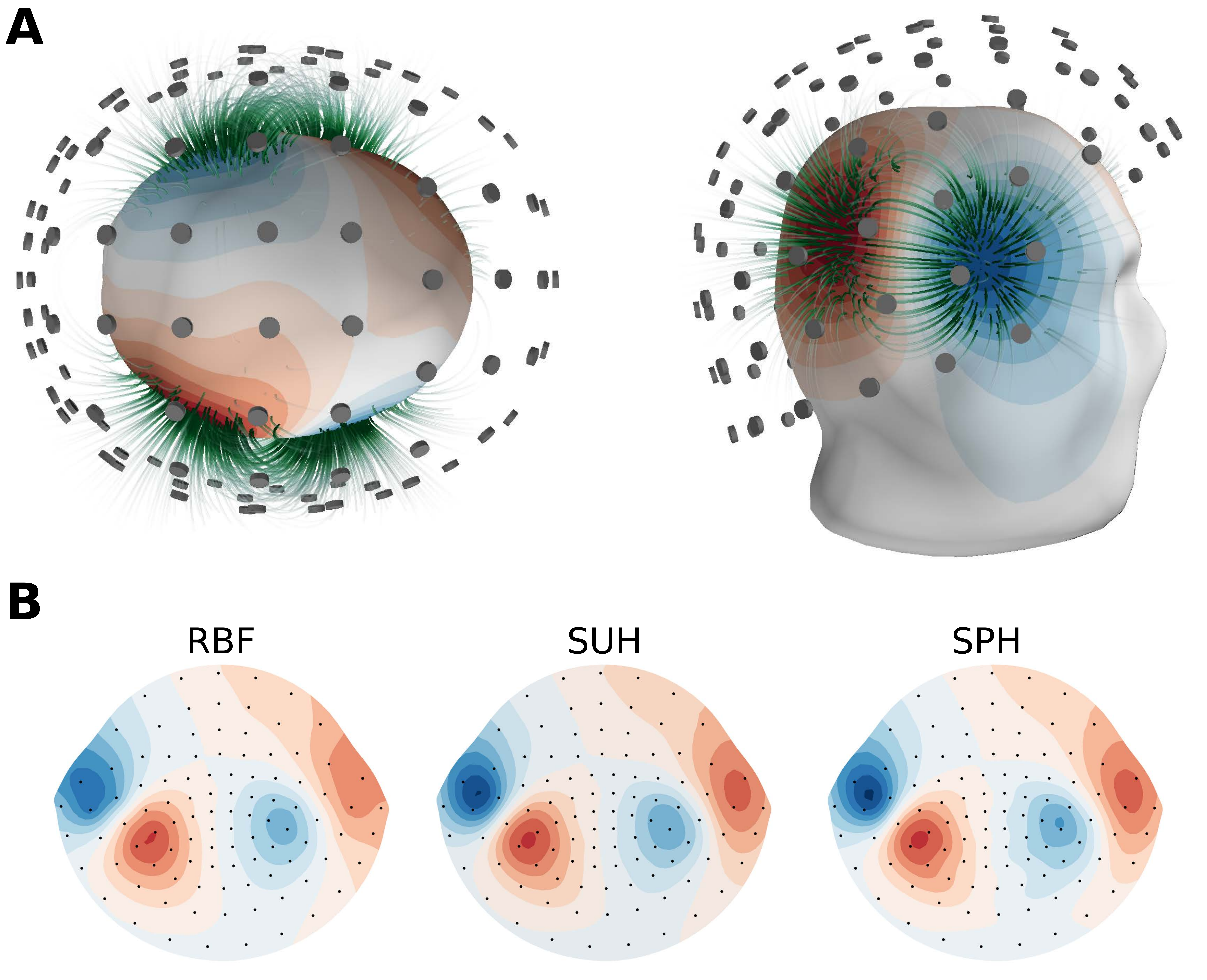}
    \caption{A: Equivalent surface current representation of an auditory evoked field measured with MEG. The stream function on the subject's scalp surface (red--blue colors) is shown in top and side views, while stream lines represent the magnetic field reconstruction. The MEG sensors are shown as grey discs. B: Topographic magnetic field map on the sensor array surface, flattened for visualization purposes. Shown is an interpolation of the measured field on the flattened 2D surface using radial-basis functions (RBF; left) as well as the surface harmonics-based (SUH; middle) and the multipole series-based (SPH; right) reconstructions. The plots have identical color scales, sensor positions are shown as black dots.}
    \label{fig:field_interp}
\end{figure}

\noindent In this example, we represent a measured magnetic field using an equivalent surface current density. We use the equivalent current density to inter- and extrapolate the magnetic field in source-free space. Specifically, we apply the equivalent surface current representation to magnetoencephalography (MEG), in which the magnetic field produced by neural currents in the brain is measured using sensors positioned around the head. We use MEG data from the sample dataset of the MNE-Python software \citep{gramfort_mne_2014}. The MEG data consists of measurements from 102 SQUID magnetometers sampled at 1 kHz during the presentation of repeated auditory beeps to the subject being measured. The magnetometers are oriented such that they measure the magnetic field component roughly normal to the subject's scalp surface (see Fig. \ref{fig:field_interp}A).

We use the subject's scalp surface (extracted from MR images) as the domain for the equivalent currents. Note that any surface that confines the ``real'' source currents generating the measured field would work. We use a regularized least-squares method to estimate the equivalent current distribution (corresponding to the stream function $\mathbf{s}$) that attempts to reconstruct the measurements $\mathbf{y}$:
\begin{equation}\label{eq:suh_fit}
\begin{split}
    \mathrm{minimize} \ E(\mathbf{s}) &= \mathbf{s}^{\top}(-\mathbf{L})\mathbf{s} + \lambda \|\mathbf{C}_{B_n}\mathbf{s} - \mathbf{y} \|^2\\
    &= \left(\mathbf{C}_{B_n}^\top \mathbf{C}_{B_n} + \frac{1}{\lambda} \mathbf{L}\right)^{-1} \mathbf{C}_{B_n} \mathbf{y}\,,
\end{split}
\end{equation}
where the first term measures the norm of the current density over the surface with $-\mathbf{L}$ being the negative Laplacian operator (meaning that we assume the current density to be maximally smoothly varying), and the second term represents the residual between the measurements and the surface-current reconstruction. Here, $\mathbf{C}_{B_n}$ is a coupling matrix that maps the stream function $\mathbf{s}$ to the measured magnetic field component $B_n$ at the sensor positions and $\lambda$ is a trade-off parameter to control the penalty on the residual in the reconstruction of the measurements. 

To express the equivalent surface current in a compact manner, we apply a truncated surface-harmonic basis. The number of components is chosen such that it is large enough not to affect the result. For regularization, we use $\lambda = 0.1 \times \text{max}(\mathbf{L}) / \sigma_\text{max}$, where $\sigma_\text{max}$ is the maximum eigenvalue of the matrix product $\mathbf{C}_{B_n} \mathbf{C}_{B_n}^\top$.

\begin{lstlisting}[language=Python]
scalp = MeshConductor(mesh_obj=scalpmesh,
                  basis_name='suh', N_suh=150)

# Magnetic field coupling from the scalp to the sensors
# taking dot product over correct dimensions
sensor_coupling = np.einsum('ijk,ij->ik',
                            scalp.B_coupling(sensor_pos),
                            sensor_normals)

# The matrix inversion
inv_cov = sensor_coupling.T @ sensor_coupling +
          lambda_ * -scalp.laplacian

#Estimate the stream function using least-squares
s = np.linalg.solve(inv_cov, 
                    sensor_coupling.T @ measured_field)

#Compute the field in any exterior point
interp_field = scalp.B_coupling(points) @ s
\end{lstlisting}
Finally, using the estimated equivalent current density, we can compute the field at any point outside the scalp surface. 
The estimated surface stream function and its magnetic field reconstruction surrounding the head during the first peak of the auditory evoked response ($t=0.080$--$0.090 \text{ms}$) is shown in Fig. \ref{fig:field_interp}A.

\subsection{Field interpolation using spherical harmonics}

\noindent In this example, we use the same data as in the previous example, but now we fit the data using spherical multipole components. We don't utilize the scalp surface, but instead construct a multipole series with the origin at the approximate center of the sensor array. The inner expansion of the multipole series is bounded by a sphere that fits between the scalp surface and the sensor array. If all measurements are outside the inner expansion volume, and all active sources are within the volume, the $\alpha$-coefficients will determine the entire field \cite{taulu_presentation_2005}.

We compute the fit for the inner expansion coefficients $\boldsymbol{\alpha}$ using the same regularized least-squares method as in the previous example (Eq. \ref{eq:suh_fit}). We replace the scalp surface Laplacian by the surface Laplacian on a sphere. The regularization is set to $\lambda=10^{-6} \sigma_{\text{max}}$, where $\sigma_{\text{max}}$ is the maximum eigenvalue of the matrix product $\mathbf{C}_{B_\alpha, n} \mathbf{C}_{B_\alpha, n}^\top$.

\begin{lstlisting}[language=Python]
# Inner expansion radius
R = np.min(np.linalg.norm(sensor_pos, axis=1)) - 0.02

lmax = 9  # maximum degree

# Compute magnetic field coupling to 
# alpha- and beta-coeffs at the sensor positions
Bca, Bcb = basis_fields(sensor_pos, lmax, normalization="energy", R=R)

# Take sensor orientation into account
Bca_sensors = np.einsum("ijk,ij->ik", Bca, sensor_normals)

# Compute Laplacian on a sphere
L = np.diag([l * (l + 1) for l in range(1, lmax + 1)
               for m in range(-l, l + 1)])

# The matrix inversion
inv_cov = Bca_sensors.T @ Bca_sensors + lambda_ * L

# Estimate the alpha-coeffs using least-squares
alpha = np.linalg.solve(inv_cov, Bca_sensors.T @ field)
beta = np.zeros_like(alpha)

# Compute the field in any exterior point
sphtools.field(points, alpha, beta, lmax,
               normalization="energy", R=R)
\end{lstlisting}

Having computed $\boldsymbol{\alpha}$, we can now compute the magnetic field at any point in the outer expansion volume. A comparison between the magnetic field computed using the multipole series fit and the surface harmonics fit (Section \ref{sec:suh_interp}) at the sensor array surface can be seen in Fig. \ref{fig:field_interp}B. For comparison, the Figure also shows a 2D interpolation of the sensor data using multiquadric radial-basis functions, similar to the interpolation method used for visualizing MEG data in the MNE-Python software package \citep{gramfort_mne_2014}. 

\section{Discussion}

\noindent In this work, we have presented the features and different components of the \mytexttt{bfieldtools} software. Further, we have showcased its usage by several examples, including code snippets and visualizations.


\subsection{Software}
\noindent Python has become a de-facto standard language for scientific software \citep[see, e.g.,][]{virtanen_scipy_2020}. We implemented \mytexttt{bfieldtools} in Python due to the rich open-source software ecosystem and large number of available libraries. In addition, Python allows for easy deployment of the software package across multiple platforms. Although the installation of the software depedencies is generally simple using official package installers (e.g. \mytexttt{pip}, \url{https://pip.pypa.io/en/stable/}) for the numerical solvers used in the coil optimization, the installation may include more complicated steps and vary across platforms.

We assume that the prospective users of \mytexttt{bfieldtools} are interested in understanding the details of the inner workings of the software. For that purpose, we strive to keep the software workflow straightforward and transparent by not hiding the NumPy arrays and other workings behind unnecessary layers of abstraction. For typical use, the \mytexttt{MeshConductor} class does include convenience functions and wrappers that reduce the need for explicit vertex indexing, function calls and extraneous variables. However, we also expose all intermediate and lower-level functions for advanced use and for, e.g, implementation of new functionality. 

\subsection{Numerical operations and discretization}
\noindent While \mytexttt{bfieldtools} does not include a meshing tool to create triangle surface meshes, most meshing tools used for finite-element modeling (FEM) or other physics modeling applications should produce meshes usable in \mytexttt{bfieldtools}. Typical FEM meshing rules of thumb also apply: the triangles should have small aspect ratios (preferably equilateral) and the mesh should have high enough resolution for the piecewise linear stream function to accurately represent the phenomena of interest. Narrow areas or areas close to mesh boundaries should generally have higher resolution. 

When using functions employing quadrature approximation to compute, e.g., the magnetic field coupling matrix, the user is free to choose the quadrature scheme. In typical use, we recommend using a dense mesh with a low-order quadrature scheme, e.g. the centroid scheme, rather than using a sparse mesh with a high-order quadrature scheme. Quadrature schemes with points at the face corners or edges should be avoided, as they cause numerical issues due to singularities of the integrands.

When using CVXPY for (quadratic) optimization, the numerical solver backend can be chosen freely. Available solvers for quadratic programming include, e.g., MOSEK \citep{andersen_mosek_2000}, CVXOPT \citep{andersen_cvxopt_2019} and OSQP \citep{stellato_osqp:_2019}. The examples in this work were run using MOSEK, which we have found to provide robust performance. However, MOSEK is a commercial product, and its use may thus be limited for some users, especially non-academic ones.

The solvers employed by CVXPY typically include (strict) feasibility checks in their optimization procedure. Additionally, they may specifically report which constraints are infeasible. The user thus gets immediate feedback on the physical feasibility of the design and can immediately respond, e.g., by altering the coil specification or the geometry. 

\subsection{Performance}
\noindent The examples in this paper were run on a regular workstation computer (4-core Intel Xeon E3-1230V5, 16 GiB RAM) with fairly dense meshes (2 000--12 000 vertices). The computation time of these examples was in the order of a one to a few minutes (biplanar coil example: 1 min 5 s; eddy current example: 5 min 49 s;  magnetic shielding example: 6 min 43 s). Besides stream-function optimization, the most time-consuming part is generally the inductance matrix computation. 
For the self-inductance matrix, the computation time is roughly quadratic with respect to the number of mesh vertices with the approximate relation $t \approx 5 \times 10^{-6} \times N_\text{v}^{2.06}$ s (1 000 vertices: 7 s; 10 000 vertices: 868 s).

The use of surface harmonics speeds up many numerical operations such as the stream-function optimization. In the first example of this paper, instead of using the vertex-wise stream function representation with one degree of freedom for each of the 12 442 mesh vertices, we used a truncated SUH expansion with 100 degrees of freedom, which took 0.35 s for optimization and 3.7 s for constructing the SUH basis. By contrast, when optimizing vertex-wise, the solver ran out of memory (16 GiB) after $\sim$30 minutes. When using a more reasonably decimated mesh with 3 184 vertices and 6 076 faces, the vertex-wise stream-function optimization took 118 s.

\subsection{Outlook and future developments}\label{sec:applications_and_extensions}
\noindent In addition to the physical quantities and couplings described in Section \ref{sec:stream}, there are other quantities for which linear mesh operators have been described previously, and which could also be implemented in \mytexttt{bfieldtools}. These include, e.g., torque due to a large (static) magnetic field \citep{lemdiasov_stream_2005}, temperature \citep{sanchez_direct_2015} and electric field in volume conductors \citep{koponen_coil_2017}. These quantities are useful in specific applications and fields, e.g. torque is relevant in MRI coil design, and the electric field is especially important in TMS coil design.

The development of \mytexttt{bfieldtools} is ongoing. As the software is open source, we welcome users from the community to contribute to the development. With contributions from different fields of science and engineering, the scope of the software could be widened to new areas and use-cases.

Planned future work include the development of, e.g., dedicated data structures for different types of sensors and sensor arrays. We further strive to keep improving the software documentation and ease of access. The scope of the software could be readily extended to electric volume conductor problems in the form of the boundary-element method (BEM), where the existing integral implementations in \mytexttt{bfieldtools} can be applied.

\section{Conclusion}
\noindent We presented \mytexttt{bfieldtools}, a novel open-source software package for magnetic field modeling with surface currents. The backbone of the software is the stream-function representation of surface current on a triangle mesh. As a key feature, the software implements a flexible coil-design method applicable for a wide range of fields within physics and engineering. The release of \mytexttt{bfieldtools} as open source enables access to stream-function-based physics modeling with minimal effort.

\section*{Acknowledgments}
\noindent This work has received funding from the European Union's Horizon 2020 research and innovation programme under grant agreement No. 820393 (macQsimal), the European Research Council under ERC Grant Agreement no. 678578 (HRMEG), the Swedish Cultural Foundation under grant no. 140635 (author RZ), and by the Vilho, Yrjö and Kalle Väisälä Foundation (author AM).

\section*{Data availability}
\noindent Data sharing is not applicable to this article as no new data were created or analyzed in this study.

\section*{References}

\bibliographystyle{aipauth4-1}
\bibliography{clean_refs.bib}

\end{document}